\documentstyle[12pt,epsf]{article}
\begin{document}
\newcommand{\largepspicture}[1]{\centerline{\setlength\epsfxsize{13cm}\epsfbox{#1}}}
\newcommand{\vlargepspicture}[1]{\centerline{\setlength\epsfxsize{15cm}\epsfbox{#1}}}
\newcommand{\smallpspicture}[1]{\centerline{\setlength\epsfxsize{7.5cm}\epsfbox{#1}}}
\newcommand{\tinypspicture}[1]{\centerline{\setlength\epsfxsize{7.4cm}\epsfbox{#1}}}
\newcommand{\ewhxy}[3]{\setlength{\epsfxsize}{#2}
            \setlength{\epsfysize}{#3}\epsfbox[0 20 660 580]{#1}}
\newcommand{\ewxy}[2]{\setlength{\epsfxsize}{#2}\epsfbox[10 30 640  590]{#1}}
\newcommand{\ewxynarrow}[2]{\setlength{\epsfxsize}{#2}\epsfbox[10 30 560 590]{#1}}
\newcommand{\ewxyvnarrow}[2]{\setlength{\epsfxsize}{#2}\epsfbox[10 30 520 590]{#1}}
\newcommand{\ewxywide}[2]{\setlength{\epsfxsize}{#2}\epsfbox[0 20 380 590]{#1}}
\newcommand{\err}[2]{\raisebox{0.08em}{\scriptsize{$\hspace{-0.8em}\begin{array}{@{}l@{}}
                     \plus\makebox[0.55em][r]{#1}\\[-0.15em]
                     \minus\makebox[0.55em][r]{#2}
                     \end{array}$}}}
\newcommand{\plus}{\makebox[15pt][c]{$+$}}
\newcommand{\minus}{\makebox[15pt][c]{$-$}}
\newcommand{\mpr}{\frac{m_{\pi}}{m_{\rho}}}
\newcommand{\beq}{\begin{equation}}
\newcommand{\eeq}{\end{equation}}
\newcommand{\ZP}{Ztschrft.\ f.\ Physik}
\newcommand{\NP}{Nucl.~Phys. }
\newcommand{\PR}{Phys.\ Rev.~D }
\newcommand{\PRL}{Phys. Rev. Lett.}
\newcommand{\NPP}{Nucl. Phys. {\bf B} (Proc.\ Suppl.\ )}
\newcommand{\PL}{Phys.\ Lett.}
\renewcommand{\textfraction}{0}
\renewcommand{\topfraction}{1}
%
\title{\vskip -3cm \makebox[9.5cm]{}{\large WUB 96-27} \\
\makebox[9.5cm]{}{\large HLRZ 53/96} \\ 
\vspace{3cm}
Evaluating Sea Quark Contributions to Flavour-Singlet Operators 
       in Lattice QCD}
\author{SESAM-Collaboration: \\ N.~Eicker$^{\rm a}$, U.~Gl\"assner$^{\rm b}$, S.~G\"usken$^{\rm b}$, 
H.~Hoeber$^{\rm a}$\\ Th.~Lippert$^{\rm a}$
        G.~Ritzenh\"ofer$^{\rm a}$,
        K.~Schilling$^{\rm a,b}$, G.~Siegert$^{\rm a}$\\
        A.~Spitz$^{\rm b}$, P.~Ueberholz$^{\rm b}$, and J.~Viehoff$^{\rm b}$ \\
{\rm $^a$}HLRZ c/o KFA J\"ulich, D-52425 J\"ulich\\
          and DESY, D-22603 Hamburg, Germany,\\
{\rm $^b$}Physics Department, University of Wuppertal\\  D-42097
           Wuppertal, Germany.
}       
\date{}
\maketitle
\begin{abstract}
In a full QCD lattice study with $N_f = 2$ Wilson fermions, we seek to
optimize the signals for the disconnected contributions to the matrix
element of flavour-singlet operators between nucleon states, which are
indicative for sea quark effects. We demonstrate, in form of a
fluctuation analysis to the noisy estimator technique, that -- in
order to achieve a tolerable signal to noise-ratio in full QCD -- it
is advantageous to work with a $Z_2$-noise source rather than to rely
only on gauge invariance to cancel non-gauge-invariant background.  In the
case of the $\pi$N $\sigma$-term, we find that 10 $Z_2$-noise sources
suffice on our sample ( about 150 independent QCD configurations at
$\beta = 5.6$ on $16^3\times32$ with $\kappa_{sea} = 0.157$,
equivalent to $M_{\pi}/M_{\rho} = 0.76(1)$), to achieve  decent signals and
adequate fluctuations, rather than 300 such sources as recently used
in quenched simulations.
\end{abstract}
\newpage
\clearpage
\section{Introduction}
The direct computation of full fledged hadronic matrix elements 
containing flavour-singlet operators ${\cal A} = \bar{q} \Gamma q$,
such as the $\pi$-N
$\sigma$-term, $\sigma_N$, or the singlet axial vector current
forward matrix element between nucleon states, has posed serious
problems to lattice gauge theory. As a result there is up to now only
faint evidence for sea quark effects from full QCD lattice
computations of such matrix elements\cite{MTC}.

  The practical bottleneck is given by the computation of disconnected
diagrams, i.e.  ubiquitous insertions of ${\cal A}$ into quark loops
disconnected from the valence quark graph, to say the nucleon propagator,
$P$, in the vacuum background field.  Technically, the disconnected
insertion $T = \sum_{x}[Tr \Gamma M^{-1}_{x \rightarrow x}]$ must be
calculated configurationwise, i.e. in correlation with the hadronic
propagator.  But the cost to compute all quark loops $M^{-1}_{x
\rightarrow x}$ individually grows with volume which renders
a direct evaluation of $\sum_x$ prohibitively expensive\cite{Mandula}.

In the framework of quenched QCD, two groups 
\cite{Kentucky1,Kentucky2,Fukugita},
have recently tackled the problem by applying two different variants
of the noisy estimator technique proposed some time ago for computing
the chiral condensate $\chi$\cite{Kennedy}. The trick is to induce the
extensive character of the quark-loop insertion by inverting the Dirac
operator $M$ on a source extending over the {\it entire (spatial) volume}.
This estimator technique is biassed in form of non-closed, i.e. bilocal and
therefore non-gauge-invariant contributions that need be controlled.

The Kentucky group\cite{Kentucky1,Kentucky2} pioneered the use of a stochastic
volume source with $Z_2$-noise (SETZ) to get rid of the `bilocal crossterms',
$x \rightarrow y$. On their {\it small sample} of 50 configurations at
$\beta = 6.0$, they used about 300 such noise sources per
configuration, at the expense of a substantial overhead in their
computatinal cost.

The authors of Ref.\cite{Fukugita}, on the other hand, utilize a
homogeneous volume source (VST) and appeal to the gauge fluctuations
to suppress the non-loop bias contributions via the Elitzur theorem.
This approach however adds a substantial amount of noise to the data
and may therefore require a {\it very high statistics ensemble} average
to yield a reasonable signal.

In exploratory quenched applications, where chromofield configurations
can be created at low cost, both approaches have led to rather
encouraging results on signal quality\cite{Okawa}.  But in the context
of a full QCD sea-quark computation, computer time prevents us at
present from doing very high statistics sampling of gauge fields. So
the question is whether effects from disconnected diagrams are
accessible at all in small/medium size sampling of full QCD.

This motivates us to look more deeply into the systematics of the
stochastic estimator technique (SET) on actual full QCD configurations
with Wilson fermions.  To be specific, we will proceed by studying the
quality (size and fluctuations) of the signal from disconnected
contributions to the scalar density.  The strategy in applying
estimator techniques should be such that -- with minimal overhead --
both systematic bias (from remaining nonlocal pollution) and
fluctuations of the signal should become negligible compared to the
statistical accuracy attainable from the size of the field sample.  

On the basis of the current statistics from our hybrid Monte Carlo
pro\-duc\-tion\cite{Melbourne,Massen} we will find that it is realistic to
exploit the stochastic estimator technique for the study of quark loop
effects in full QCD.

\section{Asymptotic Expectations}
In order to determine the disconnected part of $\sigma_N$ one needs
to calculate the expectation value $\langle P\; Tr(M^{-1}) \rangle$,
where $P$ denotes the (momentum zero)
proton correlation function and $M$ is the fermion matrix.

With VST  one calculates
$\sum_{i,j}M^{-1}_{i,j}(C)$ on each gauge configuration $C$ by solving
\begin{equation}
M(C) X = h\quad , \label{Linear}
\end{equation}
 where $h$ is a volume source vector with
components  $h_i=1$. The average over
the gauge configurations
\begin{equation}
A = \frac{1}{N_C} \sum_{C=1}^{N_C} P(C)\sum_{i,j}M^{-1}_{i,j}(C)
\end{equation}
can be separated into local and non local contributions 
\begin{equation}
 A =   
\frac{1}{N_C} \sum_{C=1}^{N_C} P(C)\sum_{i} M^{-1}_{i,i}(C) + 
\frac{1}{N_C} \sum_{C=1}^{N_C} P(C)\sum_{i \ne j} M^{-1}_{i,j}(C)
\; .
\label{volume_source}
\end{equation}
As the latter are  not gauge invariant they vanish (only)
in the large $N_C$ limit:
\begin{equation}
\lim _{N_C \rightarrow \infty} A = \langle P Tr(M^{-1})\rangle \; .
\end{equation}

In the stochastic estimator technique, one uses (complex) random source vectors
with the property
\begin{equation}
\lim _{N_E \rightarrow \infty} \frac{1}{N_E}
\sum_{E=1}^{N_E} \eta^{*}_i(E,C) \eta_j(E,C) = \delta_{i,j} \label{delta_cond} 
\end{equation}
and computes the quantity $ \eta^{\dagger}(E,C) M^{-1}(C) \eta(E,C)$
repeatedly ($N_E$ times) on each configuration.

  It appears natural to choose $\eta(E,C)$ according to a Gaussian
distribution (SETG) \cite{Kennedy}. A decomposition similar to
eq.(\ref{volume_source}) then yields
\begin{eqnarray}
\lefteqn{\frac{1}{N_C} \sum_{C=1}^{N_C} \frac{1}{N_E}\sum_{E=1}^{N_E}
\eta^{\dagger}(E,C)M^{-1}(C) \eta(E,C)P(C)  
 = } \nonumber \\ 
&&\frac{1}{N_C} \sum_{C=1}^{N_C} 
\left\{ \left[ \sum_{i} M^{-1}_{i,i}(C) +   
 \bar{T}_{on}(N_E,C) + \bar{T}_{off}(N_E,C) \right]P(C)\right\} 
\nonumber \label{gauss_source} \; ,\\ 
\end{eqnarray}
where
\begin{eqnarray}
\bar{T}_{on}(N_E,C) &=&
\frac{1}{N_E} \sum_{E=1}^{N_E} 
\sum_{i} \left[ \eta_i^{*}(C,E)\eta_i(C,E) -1 \right] M^{-1}_{i,i}(C) 
\nonumber \\
\bar{T}_{off}(N_E,C) &=&
\frac{1}{N_E} \sum_{E=1}^{N_E} 
\sum_{i \neq j } \eta_i^{*}(C,E)\eta_j(C,E))M^{-1}_{i,j}(C) \;.
\end{eqnarray}
Unfortunately the term $\bar{T}_{on}(N_E,C)$ on the right hand side of
eq.(\ref{gauss_source}) introduces a gauge invariant bias, whose
suppression definitely requires the number of estimates per
configuration to be in the asymptotic regime, where eq.\ref{delta_cond} becomes
valid.  

This dangerous bias is removed from the beginning, if one
samples the components of the random vector $\eta(E,C)$ according to a
$Z_2$ distribution. In this case, the relation $\eta_i^{*}(E,C) \eta_i(E,C) =
1$ holds for any single estimate $E$, and $T_{on} $ vanishes on each
gauge configuration. 

For the remaining nonlocal bias, $\bar{T}_{off}(N_E,C)$, SETZ combines
 two additive suppression mechanisms: (a) the gauge fluctuations
 cancel them as non gauge invariant objects on a sufficiently large
 sample of gauge fields, even for $N_E= 1$; (b) the $Z_2$-noise kills
 them in the large $N_E$ limit, even on a single gauge configuration.

In order to understand  the efficiency of the competing  methods 
for rendering good signals in more   detail,
we consider the variance $\sigma^2$ of $\langle P Tr(M^{-1}) \rangle$
in each case. Asymptotically one finds
\begin{equation}
\sigma^2 = \sigma^2_{gauge}  
 + \left\{ \begin{array}{ll}
\frac{1}{N_E} \langle \sigma^2_{off} P^2 \rangle 
& \mbox{for SETZ}\\
\\
\frac{1}{N_E} \langle (\sigma^2_{on} + \sigma^2_{off}) P^2 \rangle 
& \mbox{for SETG} \\
\\
\langle (\sum_{i \neq j} M_{i,j}^{-1})^2 P^2 \rangle 
& \mbox{for VST}
\end{array} \right\} 
 \quad + 2\; \mbox{COV} \quad .
\label{Sigma_full}
\end{equation}
%
$\sigma^2_{gauge}$ is the variance calculated with the exact
value of $Tr M^{-1}$ on each configuration, and $\sigma^2_{off}$ and
$\sigma^2_{on}$ are the variances due to the distribution of $T_{off}$
and $T_{on}$ within the  process of stochastic estimation.
The cornered brackets
stand for the average over gauge configurations.
The abbreviation COV in eq.\ref{Sigma_full} reads in its full length
\begin{equation}
COV = cov \left\{
\begin{array}{ll}
(P\;Tr(M^{-1}),P \bar{T}_{off}) & \mbox{for SETZ} \\ 
& \\
(P\;Tr(M^{-1}),P (\bar{T}_{on} +\bar{T}_{off})) & \mbox{for SETG} \\
& \\
(P\;Tr(M^{-1}),P\sum_{i \neq j} M_{i,j}^{-1})& \mbox{for VST}
\end{array}
\right. \; .
\label{Covar_long}
\end{equation}
 These formulae\footnote{A similar analysis has been carried out in 
\cite{Japan_pipi} for a modified version of VST, applied to 
the calculation of the $\pi\pi$ scattering length.}
show how the signals will be blurred by the
fluctuations of the estimates $Tr(M^{-1}(C))$.  Note that the terms in
the large curly brackets of eq.\ref{Sigma_full} are all gauge invariant.
Hence they survive
even in the limit of an infinite sample of gauge configurations 
unless suppressed otherwise : there is a way to fight them within
SETZ and SETG according to a $1/N_E$ suppression\footnote{The terms
depicted by COV behave like $\sqrt{1/N_E}$ for SETZ and SETG. In case
of SETG COV contains a gauge invariant part due to $\bar{T}_{on}$, which
can be removed only in the limit $N_E \rightarrow \infty$. All other
terms in COV vanish for $N_C \rightarrow \infty$, as they are not
gauge invariant.},
but there is no way to influence them at all within VST. \\

So far the discussion is qualitative only, as we do not know the
relative size of these additional terms in eq.\ref{Sigma_full}.
Furthermore, the calculation of $\sigma_N$ requires the
investigation of the ratio 
$\langle P(0 \rightarrow t)
Tr(M^{-1}) \rangle / \langle P(0 \rightarrow t) \rangle $ rather than
$\langle P(0 \rightarrow t) Tr(M^{-1}) \rangle $ itself. This can
entail cancellations due to additional correlations between numerator
and denominator.  Last not least we have to keep in mind that these
considerations are only valid in the asymptotic limits of the gauge
and estimator samples.  In the next section we will therefore present
a {\it numerical} study of the situation, under actual working conditions
of full QCD.

\section{Numerical Results}
Our present analysis of $\sigma^{disc}_N$ is based on 157
configurations from our ongoing Hybrid Monte Carlo run\cite{Melbourne,
Massen} with two degenerate flavours of dynamical Wilson
fermions. Here we work with $\kappa_{sea} = .157$, which amounts to a
ratio $M_{\pi}/M_{\rho} = 0.76(1)$, and to a quark mass of
$m_q \simeq 1.3 m_s$.

\begin{figure}[htb]
\epsfxsize=12.0cm
\centerline{\epsfbox{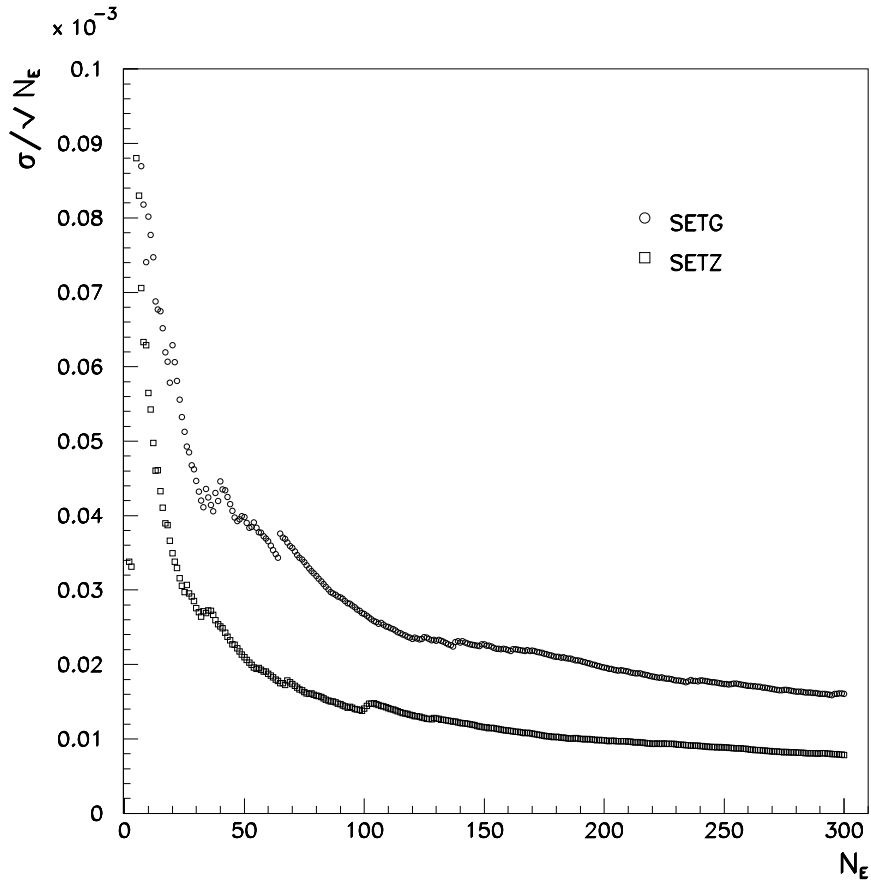}}
\vskip -0.3cm
\caption[a]{\label{Noises_error}
\it The standard error of $\chi$ for SETG and SETZ on a given configuration.}
\end{figure}

1. To set the stage, we consider first the implications of the
stochastic estimator technique on the chiral condensate, $\chi =
Tr(M^{-1})/12 V$ and compare the performance of
Gaussian and $Z_2$-noises in full QCD. In 
Fig.\ref{Noises_error} we show -- on a given configuration -- the
standard error resulting from the two methods as a function of the
number of estimates, $N_E$.
 Clearly SETZ is superior (by about a
factor two) in obtaining a good signal for $\chi$; we note that this is similar to
the related quenched situation\cite{Kentucky1}. For this reason we
pursue SETZ in the following.

2. An obvious way to economize is to properly adjust the accuracy in
solving\cite{Solver} eq.\ref{Linear}.  We therefore compute in the
next step the chiral condensate in its dependence on the inversion
accuracy $r = \;\parallel M X - h \parallel/\parallel X \parallel $.
\begin{figure}[htb]
\epsfxsize=12.0cm
\centerline{\epsfbox{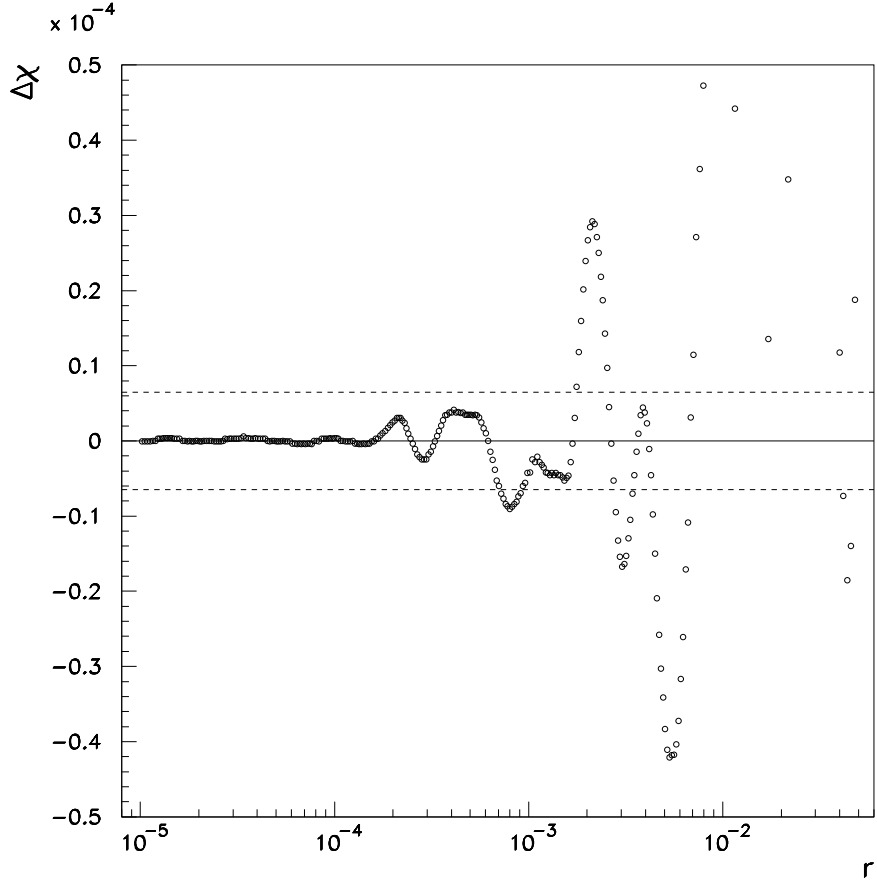}}
\vskip -0.3cm
\caption[a]{\label{Residue}
\it The difference $\Delta \chi(r) = \chi(r_{max}=10^{-5}) - \chi(r)$ on a
given configuration. The dashed horizontal lines indicate the one $\sigma$
margin from SETZ with $N_E = 300$.}
\end{figure}
Fig.\ref{Residue} shows the convergence
behaviour, $\Delta \chi = \chi (10^{-5}) - \chi (r)$. The two
horizontal lines in the plot refer to the $1 \sigma$ margin from the
stochastic noise with $N_E = 300$ estimators, on a single field
configuration. It can be seen that below $r \simeq 10^{-4}$, $\Delta
\chi$ is safely inside this margin.  It is therefore sufficient to
operate with the convergence condition $r \simeq 10^{-4}$.

\begin{figure}[htb]
\begin{center}
\noindent\parbox{13.5cm}{
\parbox{4.5cm}{\epsfxsize=4.5cm\epsfbox{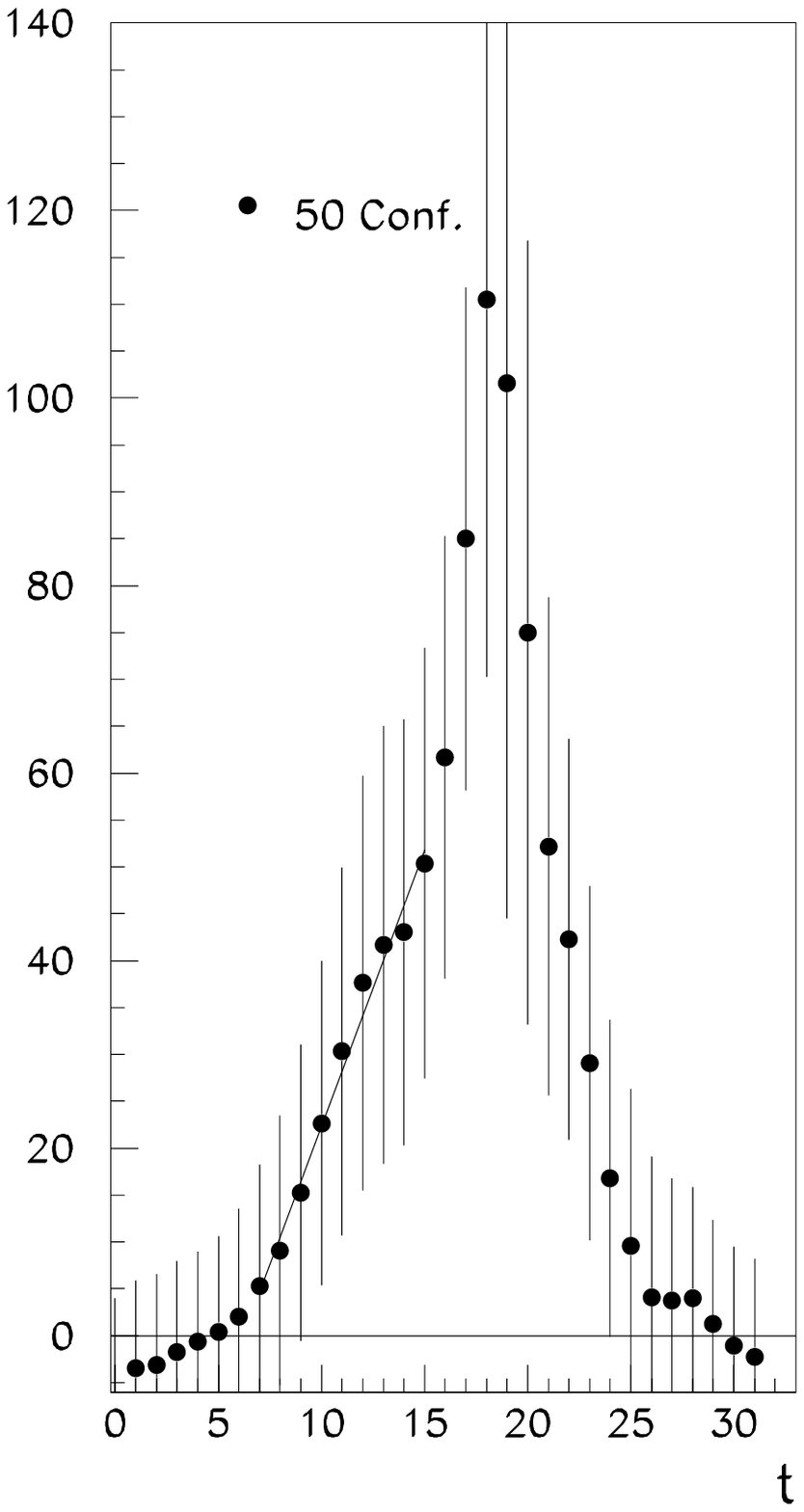}}\nolinebreak
\parbox{4.5cm}{\epsfxsize=4.5cm\epsfbox{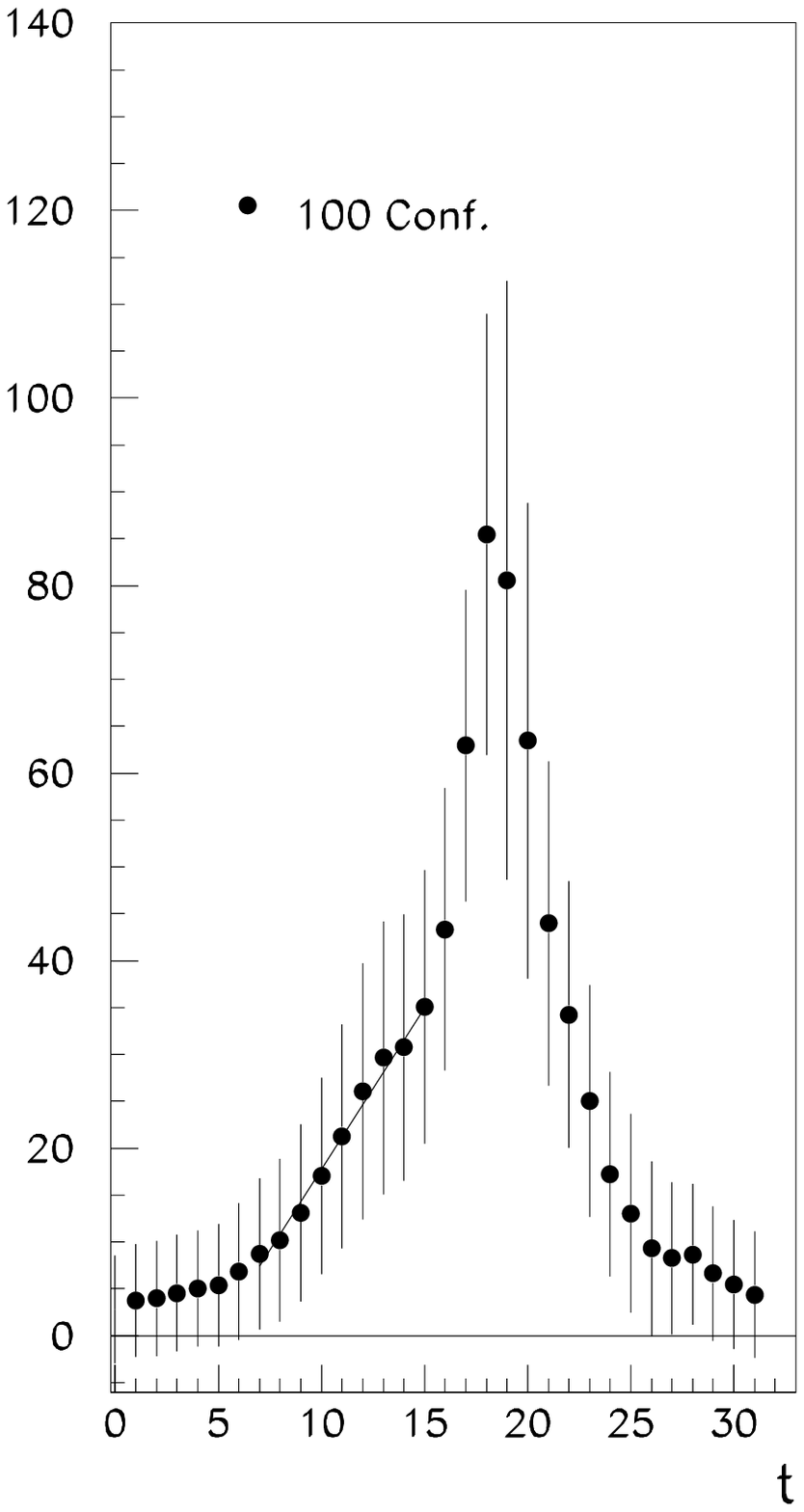}}\nolinebreak
\parbox{4.5cm}{\epsfxsize=4.5cm\epsfbox{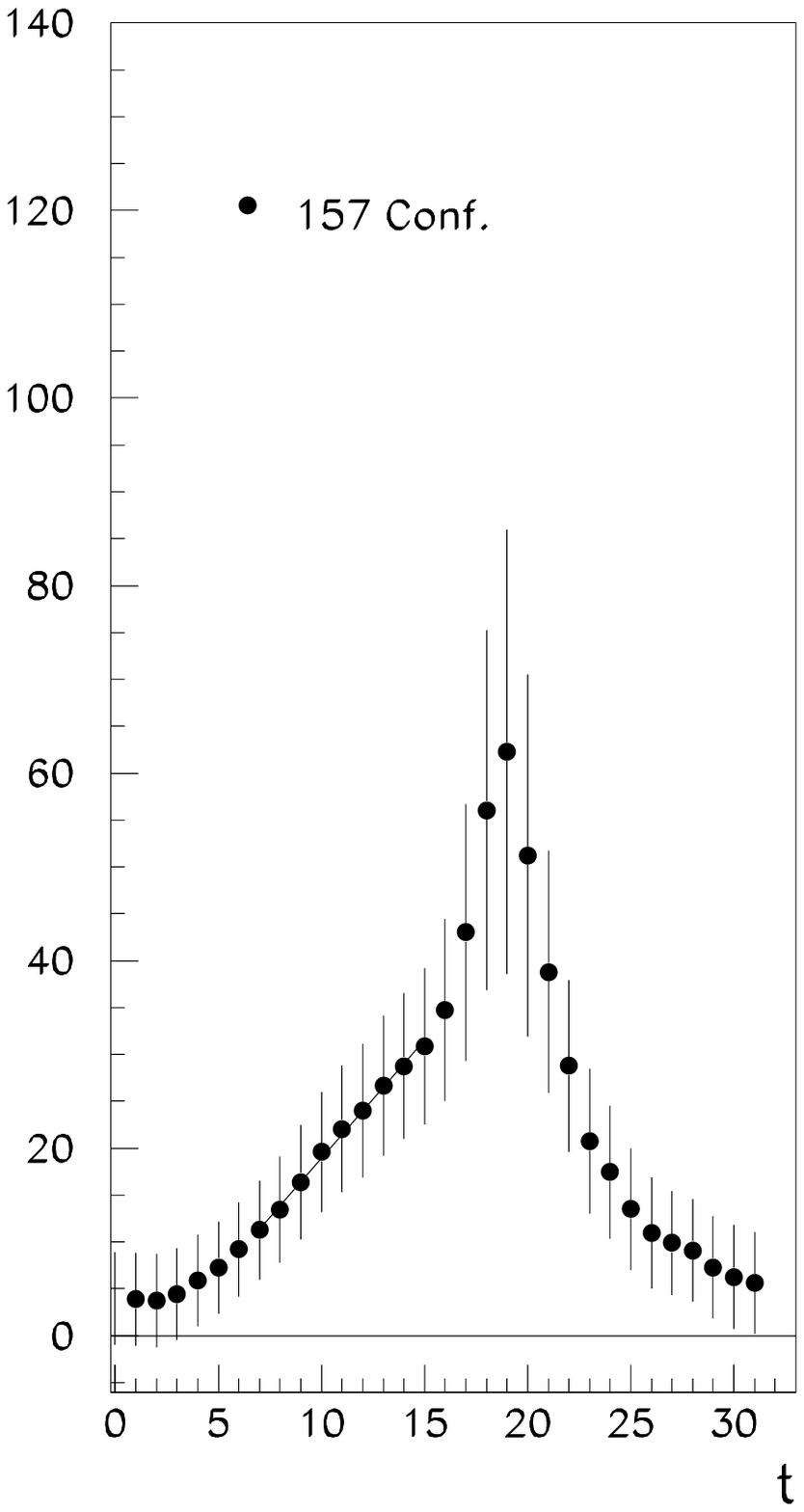}}}
\end{center}
\vskip -0.3cm
\caption[a]{\label{Sample_size}
\it $R(t)^{disc}$ measured with SETZ ($N_E=300$) on 50, 100 and 157
gau\-ge field configurations. Linear fits were performed
in the range $t = 7$ to $t =  15$.}
\end{figure}

3. The next question relates to the size of the sample of gauge
configurations required to observe a signal of the scalar density
matrix element $\langle P|\bar{q}q|P\rangle $. This 
quantity\footnote{Throughout this work we applied gauge invariant 
gaussian smearing \cite{smear} with $N=50$,$\alpha=4$ to the Proton
operator at the source.} is
extracted from
\begin{equation}
R(t)^{disc} = \frac{ \langle P(0 \rightarrow t) Tr(M^{-1}) \rangle}
             {\langle P(0 \rightarrow t)\rangle } - \langle Tr(M^{-1})
             \rangle \stackrel{t \; large}{\longrightarrow}
             \mbox{const} + t\,\langle P|\bar{q}q|P\rangle^{latt}_{disc} \; .
\end{equation}
To start we choose a large number of stochastic sources, $N_E = 300$
as proposed for quenched simulations in Ref.\cite{Kentucky2}, in order
to avoid bias from the nonlocal  pollutions to the trace.  In
Fig.\ref{Sample_size} we show how the linear rise in $R(t)^{disc}$
evolves more and more clearly as the sampling is increased from 50 to
100 and 157 gauge configurations.
\begin{figure}[htb]
\begin{center}
\noindent\parbox{13.5cm}{
\parbox{4.5cm}{\epsfxsize=4.5cm\epsfbox{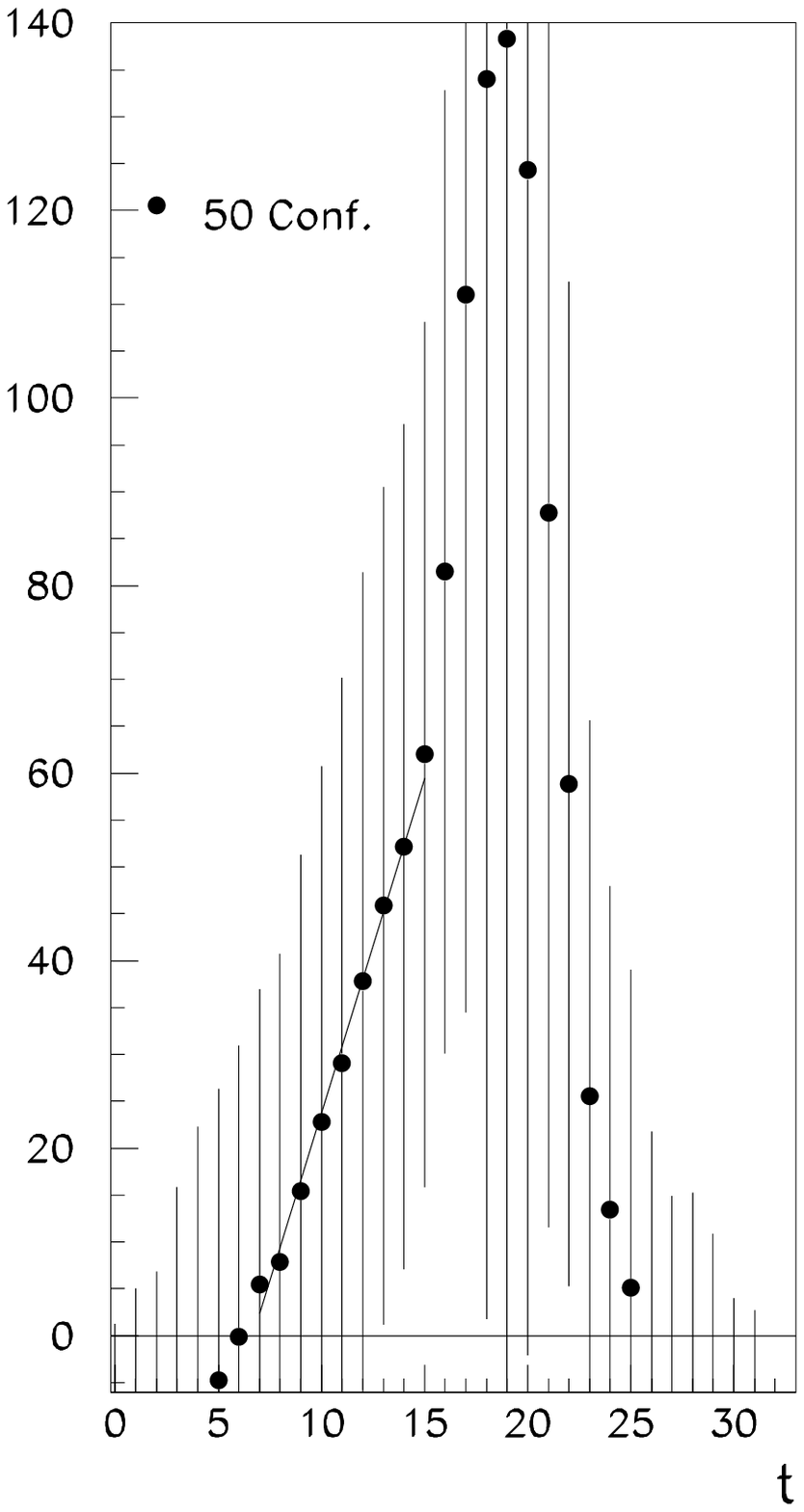}}\nolinebreak
\parbox{4.5cm}{\epsfxsize=4.5cm\epsfbox{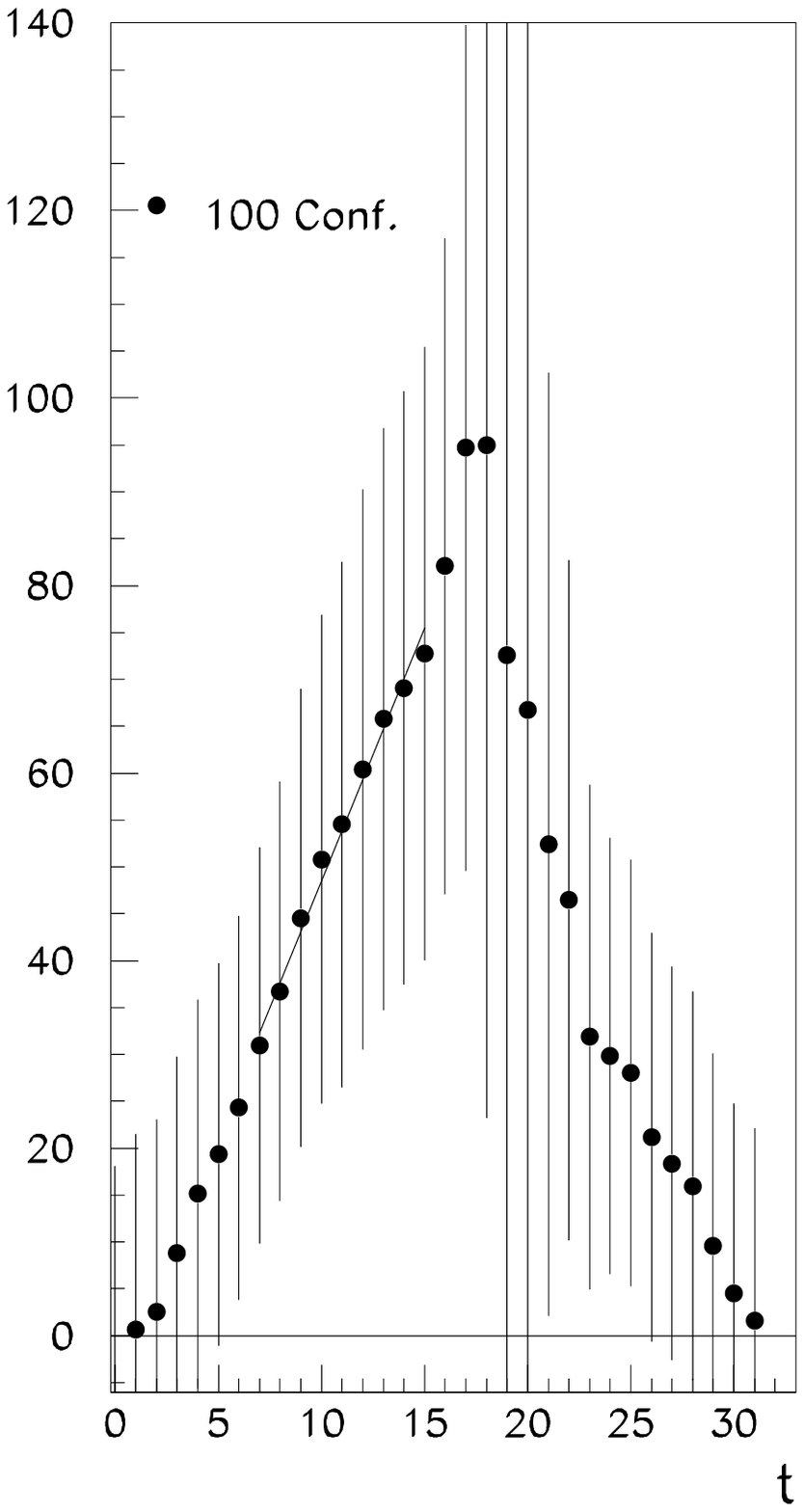}}\nolinebreak
\parbox{4.5cm}{\epsfxsize=4.5cm\epsfbox{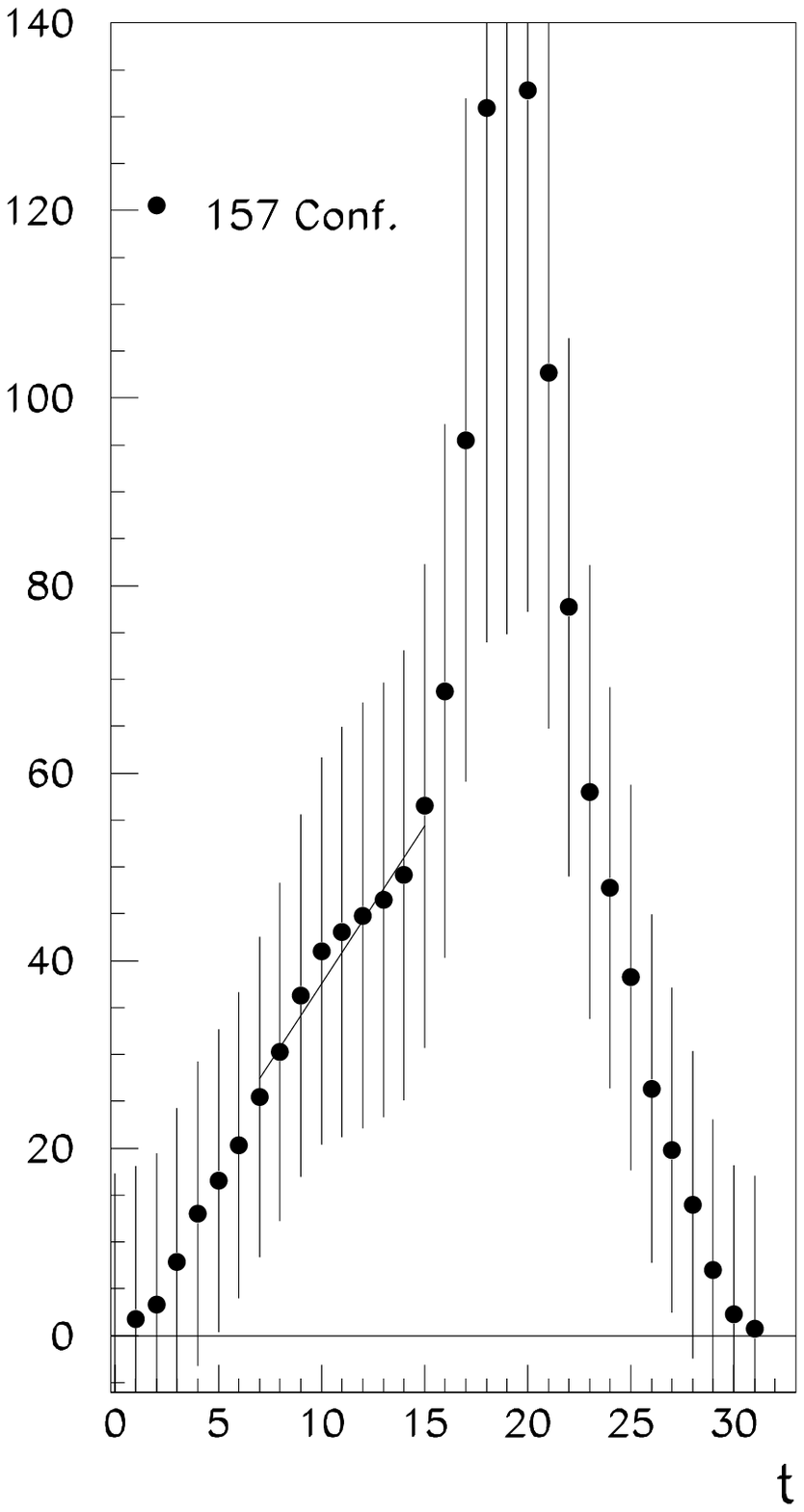}}}
\end{center}
\vskip -0.3cm
\caption[a]{\label{Sample_size_wall}
\it $R(t)^{disc}$ measured with VST on 50, 100 and 157
gauge con\-figu\-rations. Linear fits were performed
in the range $t = 7$ to $t =  15$.}
\end{figure}
The linear fits to this data yield
the values 5.89(2.37), 3.44(1.45) and 2.51(0.77) for the slope.

We conclude that it appears
mandatory to work on samples of at least 100 configurations. Under
this condition we (a) retrieve a reasonable signal and (b) find the
statistical errors on the data points to follow the expected behaviour
$1/\sqrt{N_c}$.
\begin{figure}[htb]
\begin{center}
\noindent\parbox{13.5cm}{
\parbox{6.7cm}{\epsfxsize=6.7cm\epsfbox{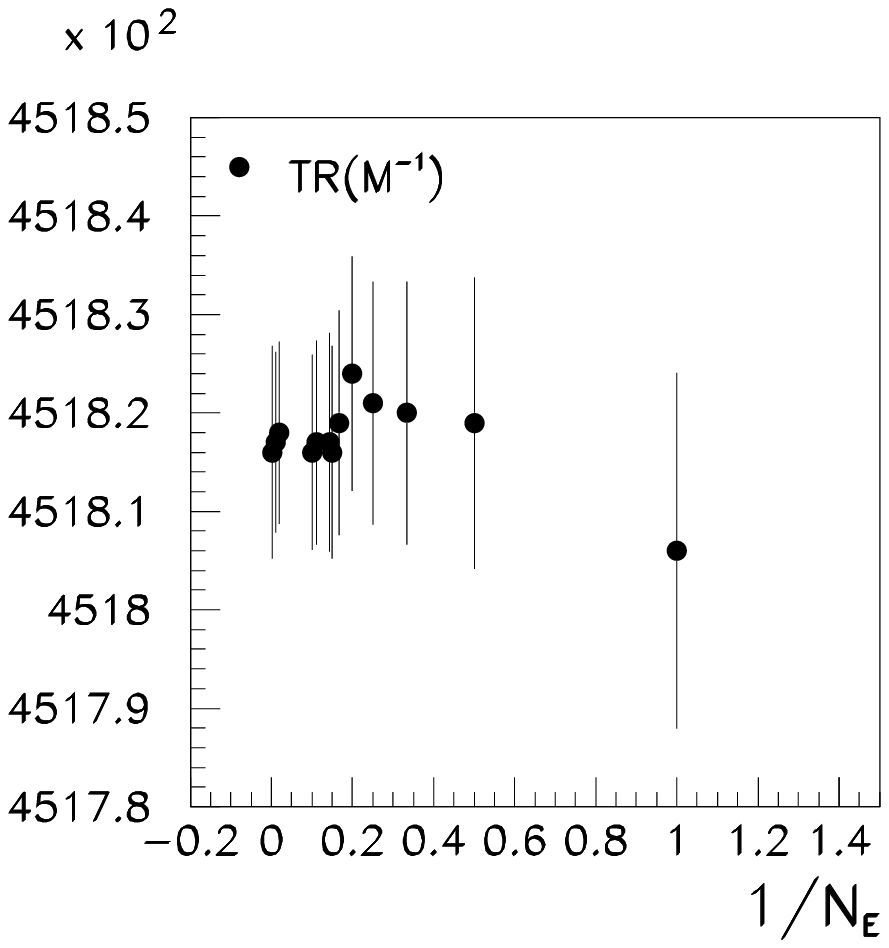}}\nolinebreak
\parbox{6.7cm}{\epsfxsize=6.7cm\epsfbox{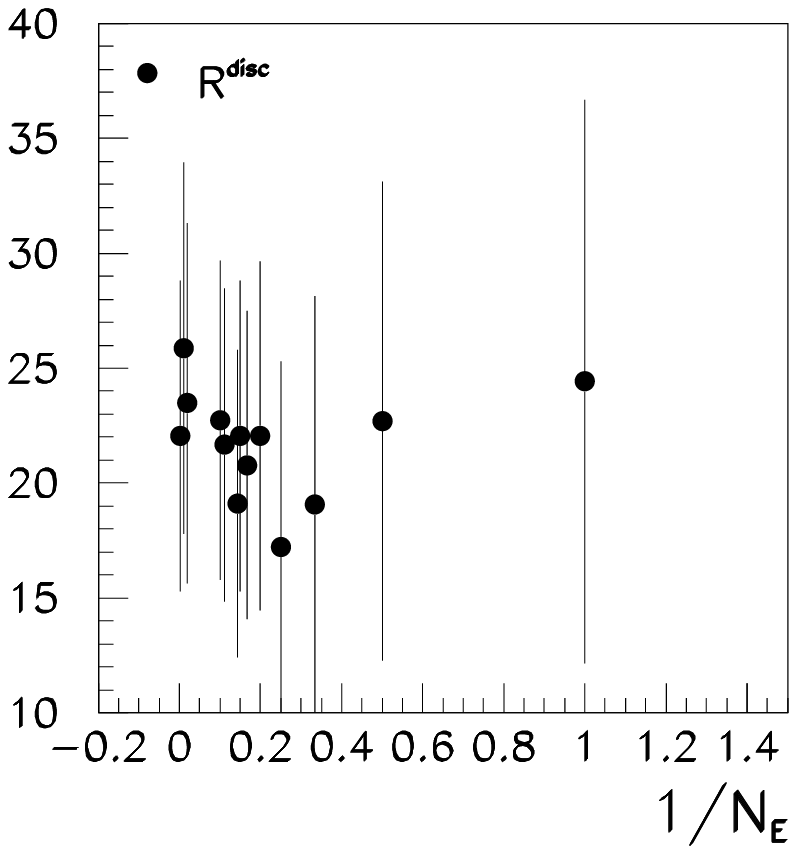}} \\
\parbox{6.7cm}{\epsfxsize=6.7cm\epsfbox{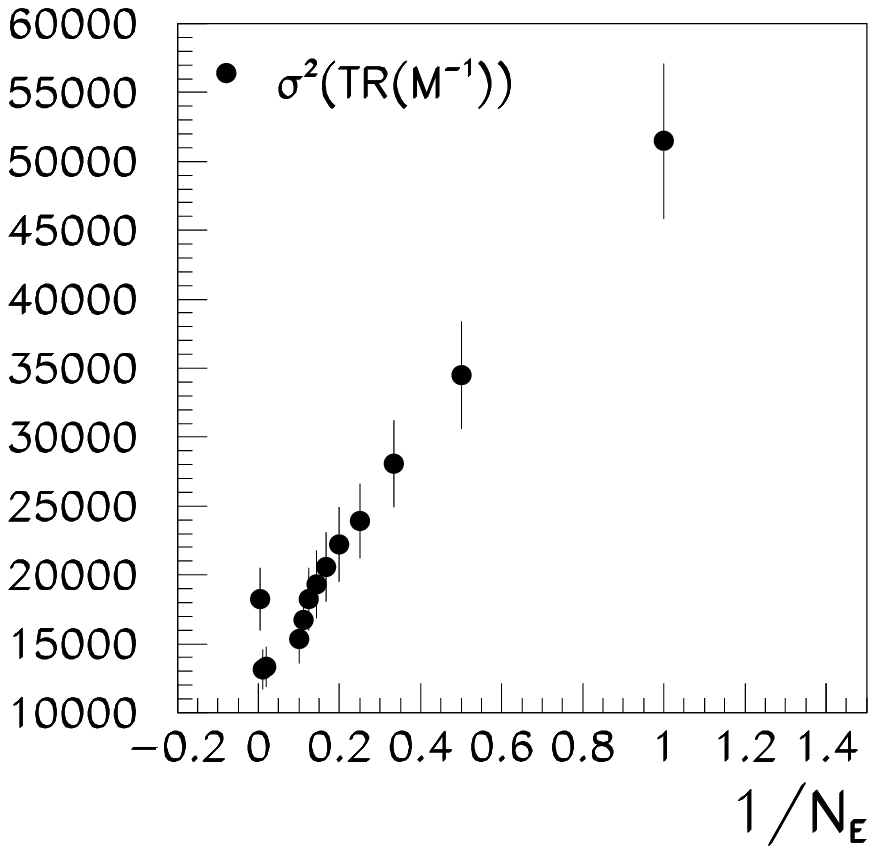}}\nolinebreak
\parbox{6.7cm}{\epsfxsize=6.7cm\epsfbox{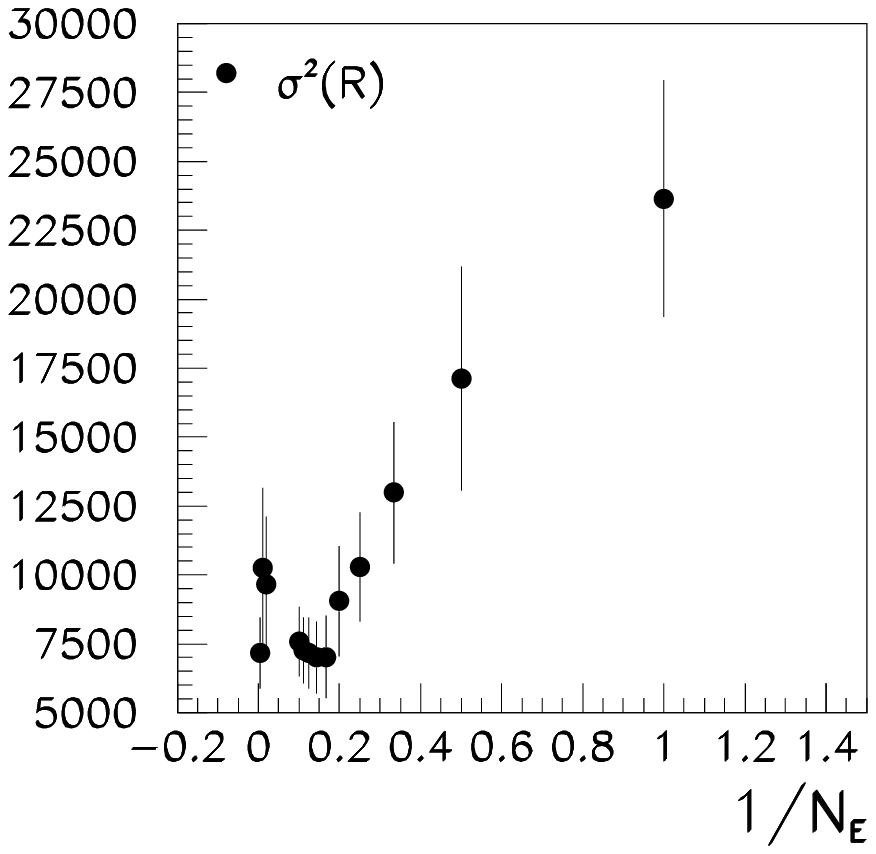}} \\
}
\end{center}
\vskip -0.3cm
\caption[a]{\label{N_E_effects}
\it $Tr M^{-1}$, $R^{disc}(t=11)$ and the corresponding variances  as a
function of $1/N_E$. The data is taken from the full set of 157
gauge configurations.}  
\end{figure}
For comparison we show in Fig.\ref{Sample_size_wall} $R(t)^{disc}$
calculated with VST on 50, 100 and 157 gauge configurations. Note that
the signal to noise ratio is much worse in this case. For the slope
we find 7.13(4.01), 5.40(2.8) and 3.50(2.20), consistent with SETZ, 
although with much larger statistical errors.
\begin{figure}[htb]
\begin{center}
\noindent\parbox{13.5cm}{
\parbox{4.5cm}{\epsfxsize=4.5cm\epsfbox{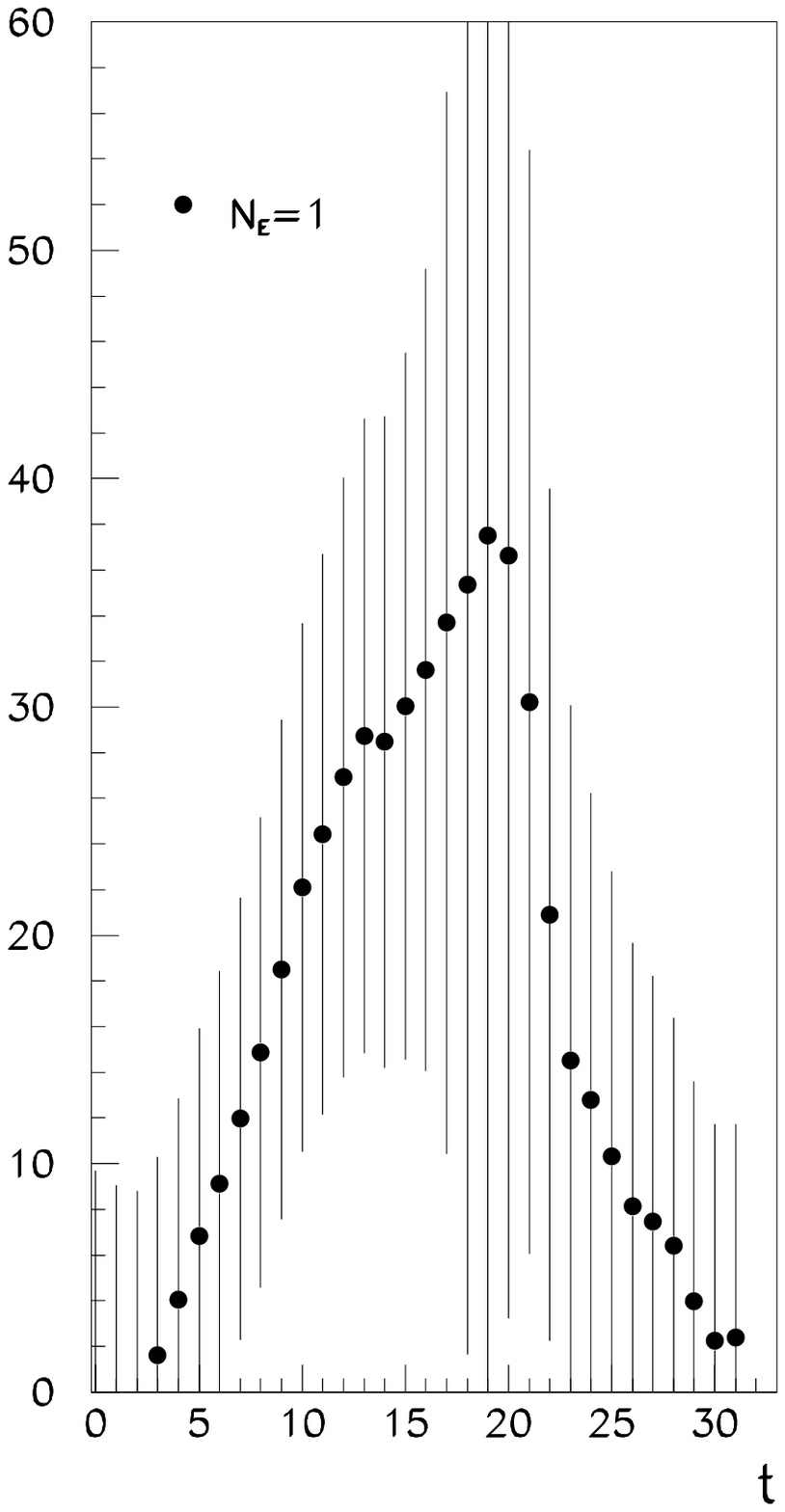}}\nolinebreak
\parbox{4.5cm}{\epsfxsize=4.5cm\epsfbox{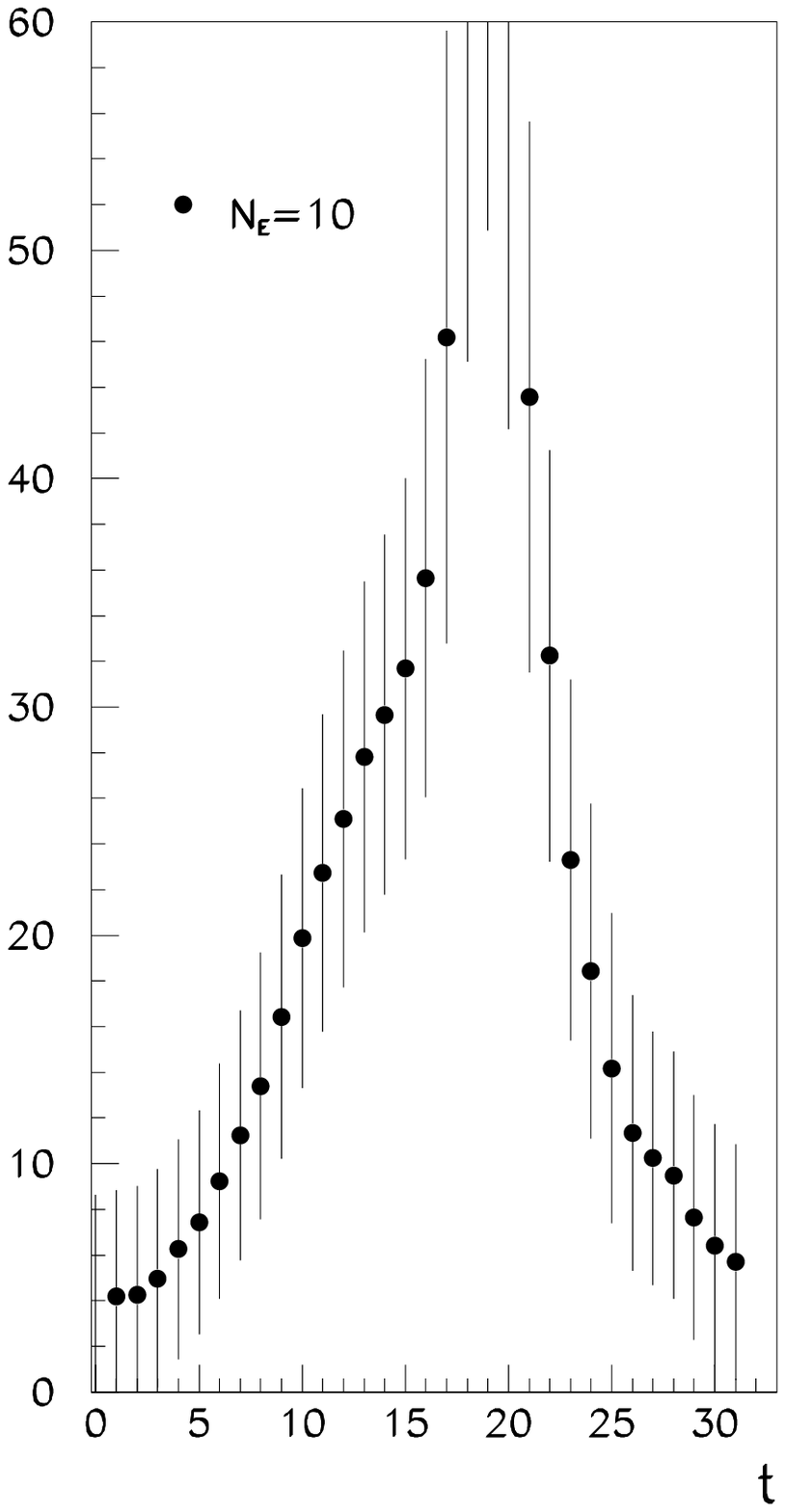}}\nolinebreak
\parbox{4.5cm}{\epsfxsize=4.5cm\epsfbox{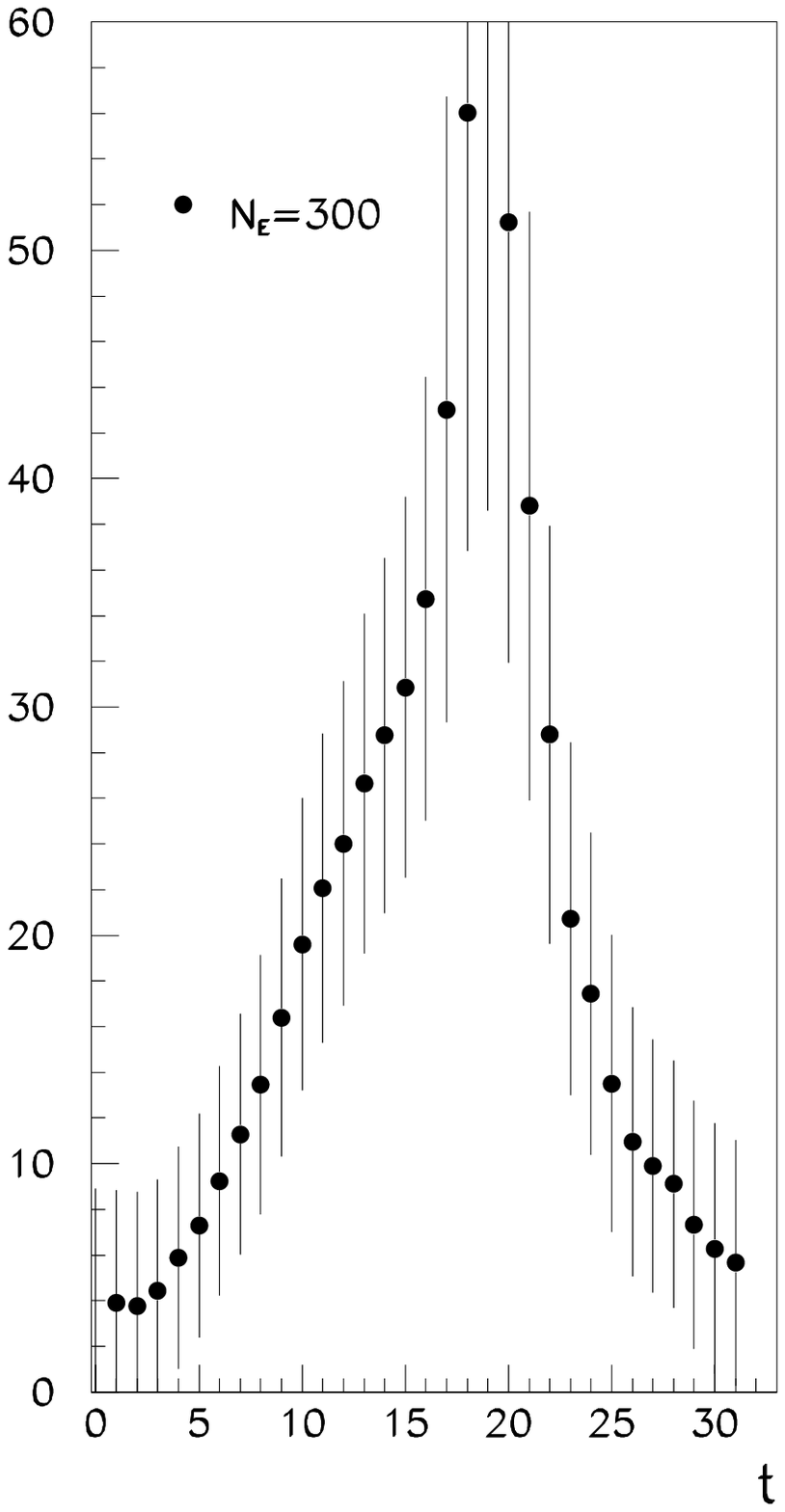}}}
\end{center}
\vskip -0.3cm
\caption[a]{\label{Full_t}
\it $R(t)^{disc}$ measured on 157 gauge configurations, with
1, 10 and 300 stochastic $Z_2$ estimates per configuration.}
\end{figure}

4. Given the signal on our present sample of 157 configurations we can
now ask the question whether and how far we can actually relax in the
number of stochastic estimates, $N_E$, without deteriorating its
quality. The starting point is the observation that the fluctuations
from SET {\it add} to the inherent fluctuations from the gauge field
sample (see eq.\ref{Sigma_full}). For economics $N_E$ should be chosen
just large enough to suppress this undesired effect.
Fig.\ref{N_E_effects} shows the response  of $Tr M^{-1}$ and
of $R^{disc}(t = 11)$ as well as the variances $\sigma^2(Tr M^{-1})$ and
$\sigma^2(R^{disc}(t = 11))$ with respect to changes in $1/N_E$.

We find that the mean values are not affected by the variation of
$N_E$ within the statistical accuracy. The variances however display 
a significant
decrease over $1 \leq N_E \leq 10$, which ends in a somewhat noisy plateau.
Obviously it does not pay to increase $N_E$ beyond a value of $10$.
For comparison we quote also the results obtained with VST on the
identical sample of gauge configurations:
$Tr M^{-1} = 451817$, $R^{disc}(t=11) = 43.07$,
$\sigma^2(Tr M^{-1}) = 121863(14582)$,
and $\sigma^2(R^{disc}(t=11)) = 75265(15073)$. 
Based on the variance we find SETZ to outperform VST
substantially. In terms of statistics we recover a gain of
a factor 2.5 to 3 from using SETZ instead of VST,  in the present
application.
The full $t$-dependence of $R^{disc}(t)$ is displayed in
Fig.\ref{Full_t}, showing again  that 10 estimates within SETZ
are sufficient to produce a reasonable signal on our sample of 157
configurations, while the signal is nil with one estimate only. On the
other hand 300 estimates are definitely (and fortunately!)
unnecessary.
\section{Conclusion and Outlook}
We have presented a full QCD study on the stochastic estimator
technique applied to the disconnected diagrams of the scalar density.
We found that it is indeed possible to achieve clean signals from only
${\cal O}(10)$ $Z_2$-noise sources with a weak convergence
requirement, $r \simeq 10^{-4}$ on the iterative solver to $Tr
M^{-1}$.  This saves a factor 30 in computer time compared to previous
applications of this technique in the quenched applications.

With as few as 157 dynamical field configurations we obtain a
reliable signal on $<P|\bar{q}q|P>_{disc.}$.
This opens the door to perform a detailed analysis of the $\pi$-N
$\sigma$ term as well as the axial vector matrix elements in the
proper setting, {\it i.e.}  without recourse to the quenched
approximation. Work along this line is in progress.

It is obvious that in full QCD simulations the use of appropriately
optimized $Z_2$-noise techniques will be of utmost importance when it
comes to the estimate of more complex quantitites in flavour-singlet
channels, where the underlying correlators contain {\it two}
disconnected fermion loops.

Exploratory quenched applications of VST to compute the $\eta '$-mass
and the annihilation diagrams in $\pi \pi$-scattering in the isospin
zero channel at threshold have attained good
signals\cite{Eta,Scattering}.  This makes us hope that medium size
sampling will lead in full QCD to reliable signals as well, once
optimized estimator techniques will be applied to grasp the loops.

{\bf Acknowledgements.}  We are grateful to DESY, DFG and KFA for
substantial computer time on their QH2 Quadrics systems at
DESY/Zeuthen, University of Bielefeld and on the Q4 and CRAY T90
systems at ZAM/KFA. Thanks to Hartmut Wittig, Markus Plagge and
Norbert Attig for their kind support. This research has been
supported by DFG (grants Schi 257/1-4 and Schi 257/3-3) and by EU
contracts SC1*-CT91-0642 and CHRX-CT92-0051.

\end{document}